\documentclass[fleqn,usenatbib]{mnras}

\usepackage{newtxtext,newtxmath}

\usepackage[T1]{fontenc}

\DeclareRobustCommand{\VAN}[3]{#2}
\let\VANthebibliography\thebibliography
\def\thebibliography{\DeclareRobustCommand{\VAN}[3]{##3}\VANthebibliography}

\usepackage{graphicx}	
\usepackage{amsmath}	
\usepackage{longtable}
\usepackage{multicol}

\title[two-component jet of GRB 191221B]{A two-component jet model for the optical plateau in the afterglow of GRB 191221B}

\author[Zhu et al.]{
Yi-Ming Zhu,$^{1,2}$
Yun Wang,$^{1,2}$
Hao Zhou,$^{1,2}$
Vladimir Lipunov,$^{3}$
David A.H. Buckley,$^{4,5,6,7}$
Pavel Balanutsa,$^{3}$
\newauthor
Zhi-Ping Jin$^{1,2}$\thanks{E-mail: jin@pmo.ac.cn}
and Da-Ming Wei$^{1,2}$\thanks{E-mail: dmwei@pmo.ac.cn}
\\
$^{1}$Key Laboratory of Dark Matter and Space Astronomy, Purple Mountain Observatory, Chinese Academy of Sciences, Nanjing 210023, China\\
$^{2}$School of Astronomy and Space Science, University of Science and Technology of China, Hefei, Anhui 230026, China\\
$^{3}$Physics Department, SAI, M. V. Lomonosov Moscow State University, Universitetsky pr., 13, Moscow 119234, Russia\\
$^{4}$South African Astronomical Observatory, PO Box 9, Observatory 7935, Cape Town, South Africa\\
$^{5}$Southern African Large Telescope, P.O. Box 9, Observatory, 7935, Cape Town, South Africa\\
$^{6}$Department of Astronomy, University of Cape Town, Private Bag X3, Rondebosch 7701, South Africa\\
$^{7}$Department of Physics, University of the Free State, PO Box 339, Bloemfontein 9300, South Africa
}

\date{Accepted XXX. Received YYY; in original form ZZZ}

\pubyear{2023}

\begin{document}
\label{firstpage}
\pagerange{\pageref{firstpage}--\pageref{lastpage}}
\maketitle

\begin{abstract}
The long gamma-ray burst GRB 191221B has abundant observations in X-ray, optical and radio bands. In the literature, the observed optical light curve of GRB 191221B displays a plateau around 0.1-day, which is rather peculiar in gamma-ray bursts. Here we performed detailed analysis of the observational data from Swift/UVOT, VLT and LCO, obtained the light curve of the multi-band afterglow of GRB 191221B. By examining optical, ultraviolet, X-ray, and radio data for this event, we demonstrate that an on-axis two-component jet model can explain the observations. Our analysis suggests that the narrow component has an initial Lorentz factor of 400 and a jet opening half-angle of $1.4^{\circ}$, while the wide component has an initial Lorentz factor of 25 and a jet opening half-angle of $2.8^{\circ}$. The narrow jet dominates the early decay, whereas the wider jet causes the optical plateau and dominates late decay. According to this model, the reason for the absence of the X-ray plateau is due to the steeper spectral index of the wide component, resulting in a less significant flux contribution from the wide jet in the X-ray bands than in the optical bands. Moreover, we have explained the inconsistency in the decay indices of the UVOT and Rc-band data around 2000 seconds using reverse shock emission.
\end{abstract}




\section{Introduction}

Gamma-ray bursts (GRBs) and their afterglows are commonly interpreted in terms of a stellar explosion with relativistic outﬂows and narrowly collimated jets \citep{PIRAN1999575}. The interaction between the jets and the external medium will lead to continuous external shock, which converts the kinetic energy of the jets into X-rays, optical, and radio multi-wavelength radiation \citep{Meszaros_1997,Sari_1998}. Understanding the properties of the jets by analyzing the afterglow is a crucial aspect of GRB research.

The afterglow of GRBs is typically modeled with a uniform jet that has a simple ``top-hat'' structure. However, for some special GRB samples, the interpretation of their afterglow behavior needs to include the consideration of the jet structure, such as considering the distribution of jet energy as a power-law \citep{2002MNRAS.332..945R,2003A&A...400..415W} or Gaussian function of angular distance from the axis \citep{Kumar_2003}.

The two-component jet model is another structured jet model that has been proposed to explain some GRBs. \citet{Berger2003-go} suggested that the observation data of GRB 030329 require a two-component explosion: the first component (a narrow jet) with higher initial Lorentz factor is responsible for the gamma-ray burst and the early optical and X-ray afterglow, while the second component (a wider jet) powers the radio afterglow and late optical emission.
\citet{Peng_2005} presented a detailed calculation for the two-component jet model and derived the flux characteristics of the two jet components at the main transition times. 
It has also been proposed in many subsequent works that for some GRBs the outflow may consist of two distinct components \citep{2004ApJ...605..300H,Oates_050802,Jin_051221A,Kamble_050401,Racusin_080319B,Filgas_080413B,Nardini_081029,Horst_130427A,Chen_160623A,Sato_190829A}.

GRB 191221B follow-up observation was triggered by the Swift satellite, and a large amount of observation data has been obtained from multiple telescopes. In this work, we present details of Swift/UVOT, the Very Large Telescope (VLT), and the Las Cumbres Observatory (LCO) observations of the afterglow of GRB 191221B and propose a two-component jet scenario to explain its optical bands, ultraviolet, X-ray, and radio behaviors.
This paper is organized as follows. In Section \ref{sec:obs} we present the multi-band observation data of GRB 191221B; in Section  \ref{sec:model}, we describe the two-component jet model and reproduce  the afterglow data with a model code; and in Section \ref{sec:dis}, we provide the conclusions and discussions. 
Throughout the paper, the cosmological constants we adopted are Hubble constant $H_0=67.4\pm0.5\,\mathrm{km}\,\mathrm{s^{-1}}\,\mathrm{Mpc^{-1}}$ and matter density $\Omega_{\rm m}=0.315\pm0.007$ \citep{2020A&A...641A...6P}.

\section{observations of GRB 191221B} \label{sec:obs}

\begin{figure*}
\includegraphics[width=1\textwidth]{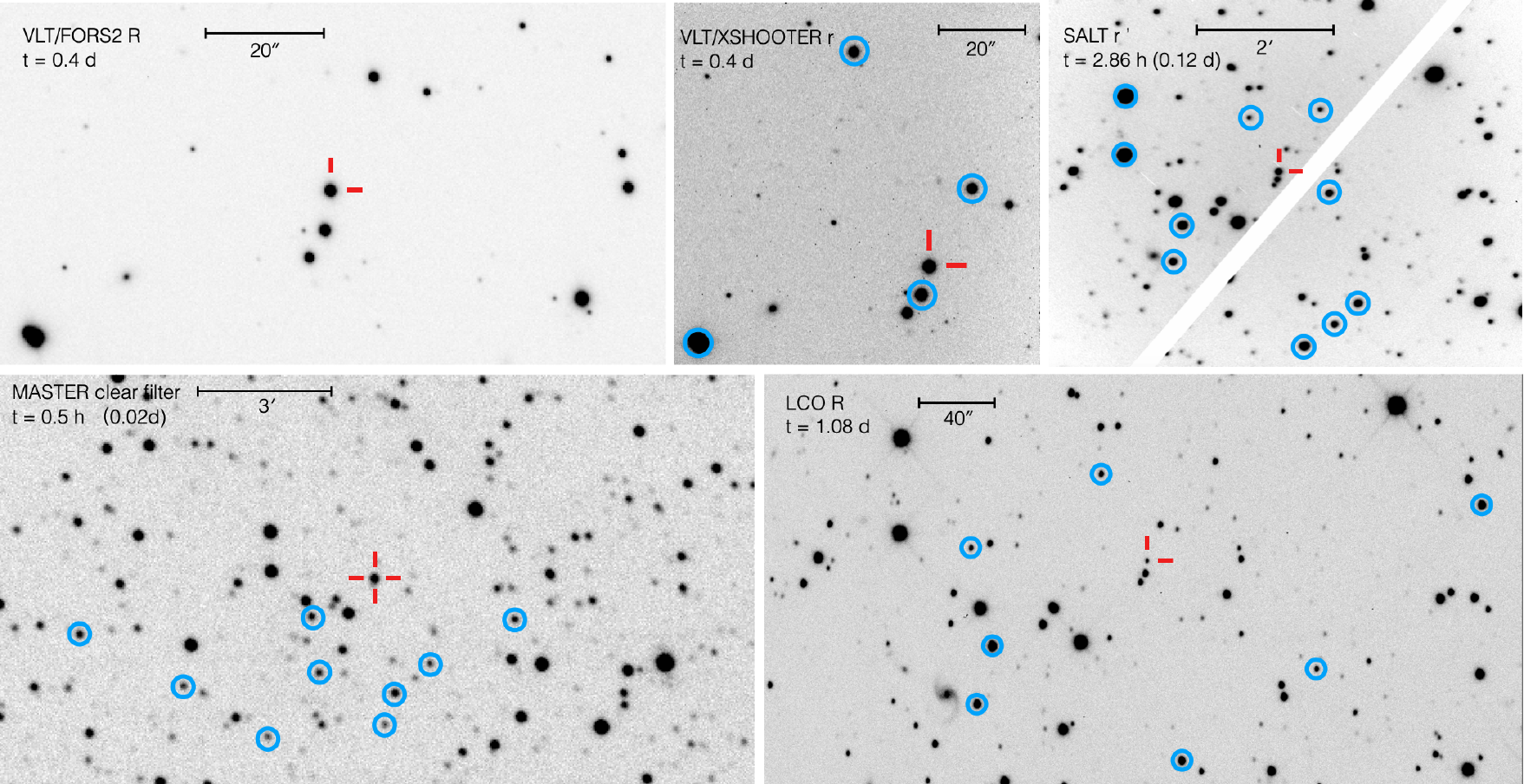}
\caption{The GRB 191221B field obtained from VLT, MASTER, SALT and LCO. The afterglow is labeled with a red cross/half cross. The blue circles indicate the standard stars used in the photometric calibration for each instrument, excluding the VLT/FORS2 image, which used a specially taken standard star calibration file.	\label{fig:allimg}}
\end{figure*}

GRB 191221B was triggered and located by the Swift Burst Alert Telescope (BAT) On 2019 December 21 at 20:39:13 UT \citep{2019GCN.26534....1L}. 
A refined analysis \citep{2019GCN.26562....1S} revealed the $T_{90}$ duration of 15-350 keV band is $48\pm 16$ s, and its photon spectrum can be fitted with a power-law index of $1.24 \pm 0.05$. 
VLT/X-shooter observed the GRB field 9.8 hours after the BAT trigger and inferred a redshift z = 1.148 from the absorption profiles in the GRB spectrum \citep{2019GCN.26553....1V}.

\cite{Buckley_2021} presented the Light curves of GRB 191221B. The MASTER clear band light curve shows a decay with a temporal slope $\alpha = 0.44$ (the afterglow data is described by $F_\nu\propto t^{-\alpha} \nu^{-\beta}$) from 0.1 days to 0.5 days, indicating that this is a evident plateau in this time period. However, there is no simultaneous plateau in the X-ray afterglow.


All MASTER telescopes have identical optical schemes and are equipped with identical sets of polarization and BVRI filters \citep{2012ExA....33..173K,2014NewA...29...65P,2017MNRAS.465.3656L,2019ARep...63..534L,2022Univ....8..271L}, enabling the derivation of a target light curve within one photometric framework \citep{2010AdAst2010E..30L,2023ApJ...943..181L,2017Natur.547..425T,2012ExA....33..173K,2010AdAst2010E..62G}.
 White light magnitudes are calculated using the equation
W = 0.2B + 0.8R \citep{2017MNRAS.465.3656L}. It is also best described by the Gaia G filter \citep{Buckley_2021}. We performed two similar photometric calibration
procedures using two different sets of reference stars from the Gaia DR2 catalogue, seven for the VWFC images, and nine for the MASTER II telescope images (see the bottom left panel of Figure \ref{fig:allimg}).

Additionally, apart from the reference stars, we have selected a large list of comparison stars with similar brightness to the object (see Figure \ref{fig:refstars}). This was done to determine the measurement error of its magnitude, considering the luminosity variation of these stars. 
For more detailed descriptions of the photometric error technique, refer to \citet{2019ARep...63..534L,2017Natur.547..425T}.

\citet{Urata2022-191221} presented the radio observations from Atacama Large Millimetre/submillimetre Array (ALMA). Eleven epochs of observations were conducted at 97.5 GHz from 0.4 days to 33.4 days, five epochs were conducted at 145 GHz from 1.4 days to 18.4 days, and two observations were conducted at 203 GHz at 1.4 days and 2.5 days.

We downloaded the VLT image data from ESO Science Archive Facility (\url{http://archive.eso.org/}). Then we reduced them using \texttt{IRAF} software \citep{1986SPIE..627..733T, 1993ASPC...52..173T}. Photometry results from the FORS2 data showed R-band magnitudes of 16.85 (AB) at 0.4 days and 20.16 (AB) at 2.4 days. The VLT/FORS2 R-band observation at 0.4-day is shown in the upper left panel of Figure \ref{fig:allimg}. VLT/X-shooter observed the afterglow of GRB 191221B in the g, r, z band at 0.4-day and 1.4-day. We reduced these X-shooter image data and measured magnitudes which are estimated from comparison to nearby stars from the Skymapper catalog \citep{Wolf_skymapper}, as shown in the upper middle panel of Figure \ref{fig:allimg}. And the upper right panel of Figure \ref{fig:allimg} presents SALT(Southern African Large Telescope) r' band acquisition image, taken at the beginning of the SALT observations at 23:30:49 UT (2.8h after the burst trigger). The afterglow magnitude for this image was also derived from the Skymapper reference stars.

LCO observed GRB 191221B from 3.5 to 4.2 hours after the GRB trigger time and detected a bright source with R and I filters \citep{2019GCN.26560....1S}. We downloaded image data from LCO's website (\url{https://lco.global}) and reduced them, then measured the magnitudes of the next three R-band observations, observed at 1.08, 2.34, and 4.10 days. These magnitudes are calibrated against several USNO-B1.0 stars \citep{Monet_2003_USNO-B} near the GRB location (see the bottom right panel of Figure \ref{fig:allimg}).



The UVOT observation data and BAT-XRT flux light curves are obtained from the Swift website (\url{https://www.swift.ac.uk}). The reduction and photometry of the UVOT data were performed using the \texttt{HEASoft} software (\url{https://heasarc.gsfc.nasa.gov/docs/software/heasoft/}). UVOT conducted extensive observations on this burst for more than ten days, but the target source could not be detected by most filters after the third day.

\begin{figure}
\includegraphics[width=0.5\columnwidth]{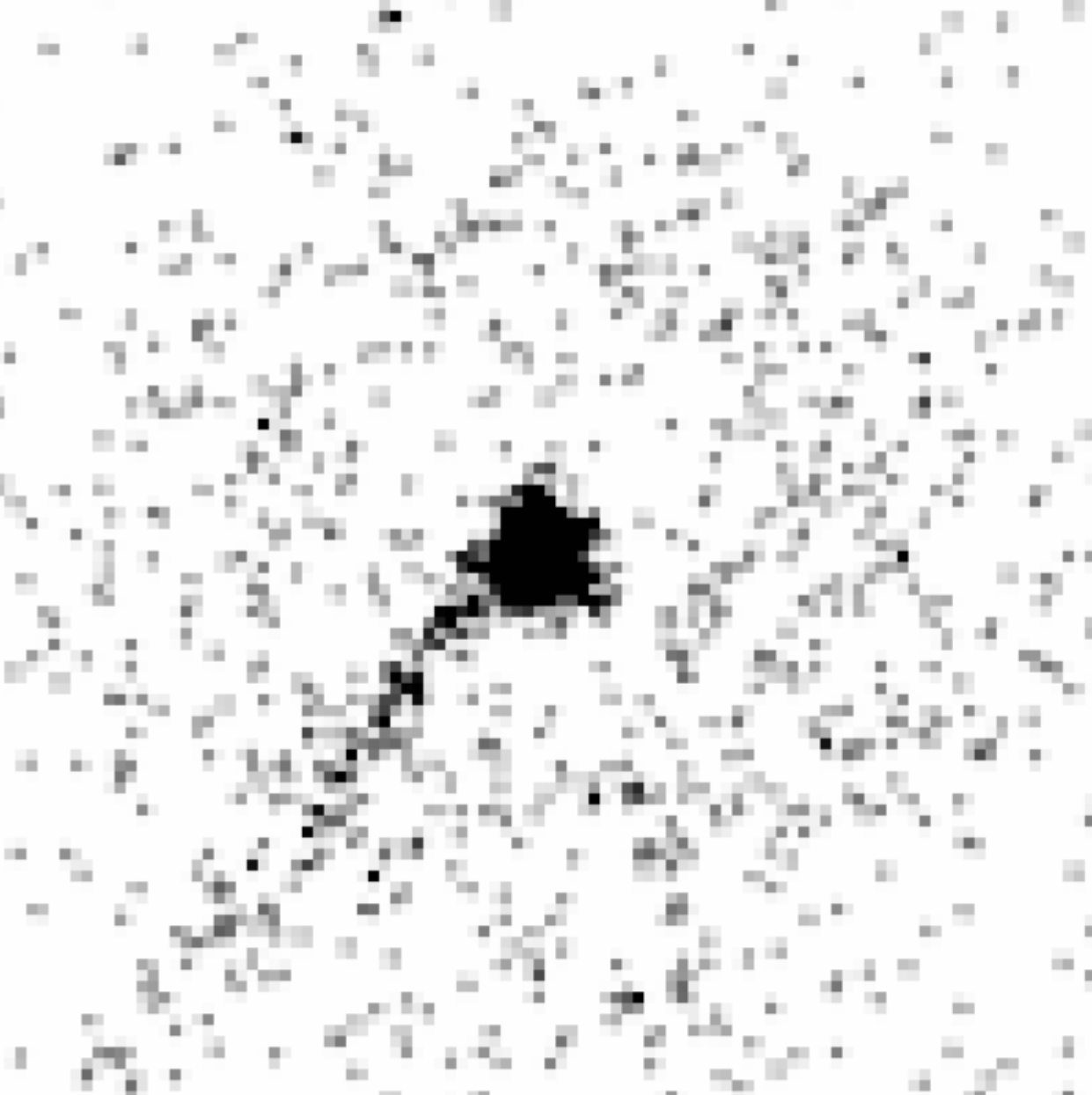}
\includegraphics[width=0.5\columnwidth]{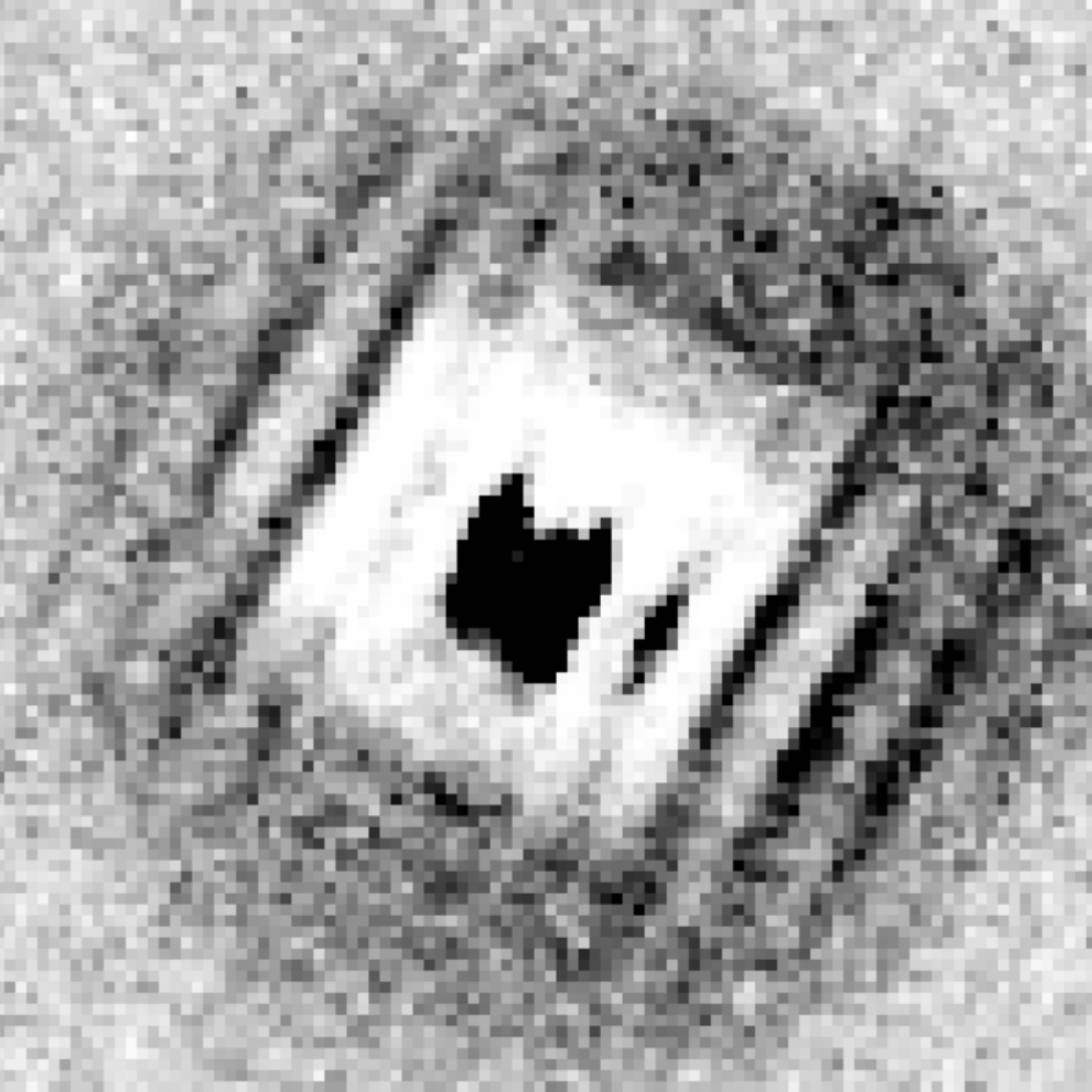}
\caption{\emph{Left panel}: The first exposure in the V-band filter of Swift/UVOT, starts at 93 seconds after the burst and lasts for 9.2 seconds. The telescope was settling at the time. \emph{Right panel}: The first exposure in the white filter of Swift/UVOT, starts at 110 seconds after the burst and lasts for 147 seconds. This image shows a saturation pattern.	\label{fig:image}}
\end{figure}

The first exposure in the V filter of Swift/UVOT started 93 seconds after the burst and lasted for 9.2 seconds. As shown in the middle panel of Figure \ref{fig:image}, the ``smear'' on the image is due to the telescope still settling at the time. After comparing with other long-exposure V-band observations, it is found that the ``smear'' in first exposure does not have a great impact on the photometric results. As \citet{Page_UVOTsettling} conclude, the settling exposures are considered to be good enough to derive photometry for the GRB afterglows. In addition, saturation patterns were observed in the images of the first four exposures in the white filter of Swift/UVOT, and the image of the first exposure is presented in the right panel of Figure \ref{fig:image}. We measured the magnitudes of these four saturated sources by using the method introduced by \citet{2023NatAs.tmp..135J}.

For a log of the observations and photometry see Table \ref{table_191221B}. Magnitudes are in the AB system and are corrected for Galactic extinction $E(B-V) = 0.067 \mathrm{mag}$ \citep{schlafly2011extinction}. The upper limit is at the $3\sigma$ confidence level.

We utilized the Small Magellan Cloud (SMC) template extinction law \citep{Li_2008_extin} to estimate the extinction of the host galaxy by fitting the afterglow spectral energy distribution (SED). The fitting procedure was performed using \texttt{emcee} Python toolkit \citep{Foreman_Mackey_2013}. The SMC extinction curve was selected as it is the preferred template for characterizing the rest-frame extinction curve in most afterglow SEDs~\citep{2015A&A...579A..74J}. The fitting yields a small visual extinction $A_V = 0.04^{+0.05}_{-0.03}$, with reduced $\chi^2 / \text{dof}=10.06/6$. This host galaxy extinction is taken into account in the subsequent figures and analysis.

\begin{figure*}
\includegraphics[width=1\textwidth]{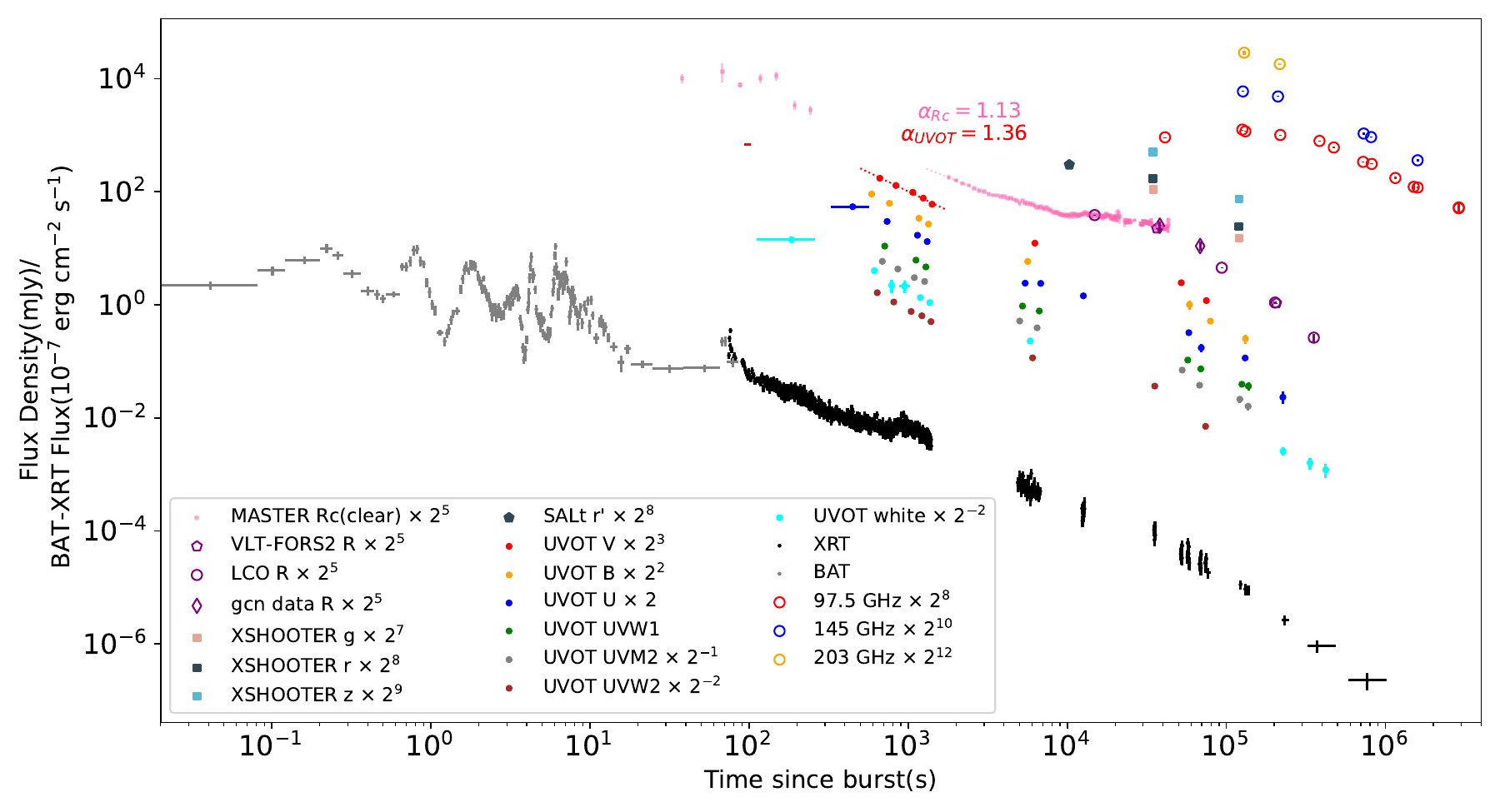}
\caption{Light curves of GRB 191221B. Optical and ultraviolet data are corrected for Galactic extinction $E(B-V) = 0.067 \mathrm{mag}$ \citep{schlafly2011extinction} and the fitted host galaxy extinction $A_V = 0.04 \mathrm{mag}$. The BAT and XRT data are presented as 0.3-10 keV fluxes. The MASTER observation data are obtained from \citet{Buckley_2021}. The radio data of ALMA observations are obtained from \citet{Urata2022-191221}.  \label{fig:lc}}
\end{figure*}

Figure \ref{fig:lc} shows the multi-wavelength light curves of the afterglow of GRB 191221B. The MASTER clear filter has a passband similar of the Gaia G filter (623nm), for which we adopt the effective wavelength appropriate to Rc-band. R-band data collected from GCN, including R=16.7 at 0.44-day \citep{2019GCN.26552....1G} and R=17.6 at 0.79-day \citep{2019GCN.26561....1G}, was also included in the figure. The BAT and XRT fluxes are on 0.3-10 keV. 

The light curve in the optical band exhibits five distinct decay phases: (A) the early decay captured by multi-bands of UVOT ($300 \mathrm{~s} \leq t-t_0 \leq 1500 \mathrm{~s}$); (B) the early decay of MASTER clear-band ($1500 \mathrm{~s} < t-t_0 \leq 3500 \mathrm{~s}$); (C) a shallow decay of MASTER clear-band until $10^4\mathrm{~s}$ ($3500 \mathrm{~s} < t-t_0 \leq 10000 \mathrm{~s}$); (D) the relatively flat decay that is considered to be a plateau ($12000 \mathrm{~s} \leq t-t_0 \leq 45000 \mathrm{~s}$); (E) the rapid decay thereafter ($> 45000 \mathrm{~s}$). To analyse the temporal behavior, we fitted a power-law function to the light curves within these five time intervals. For the UVOT light curves, we conducted a joint fit across all six bands to determine the decay indices. The fitting results are listed in Table \ref{table_fit}. The reduced chi-squares with degrees of freedom were adopted to assess the goodness of fit.

\begin{figure}
\includegraphics[width=1\columnwidth]{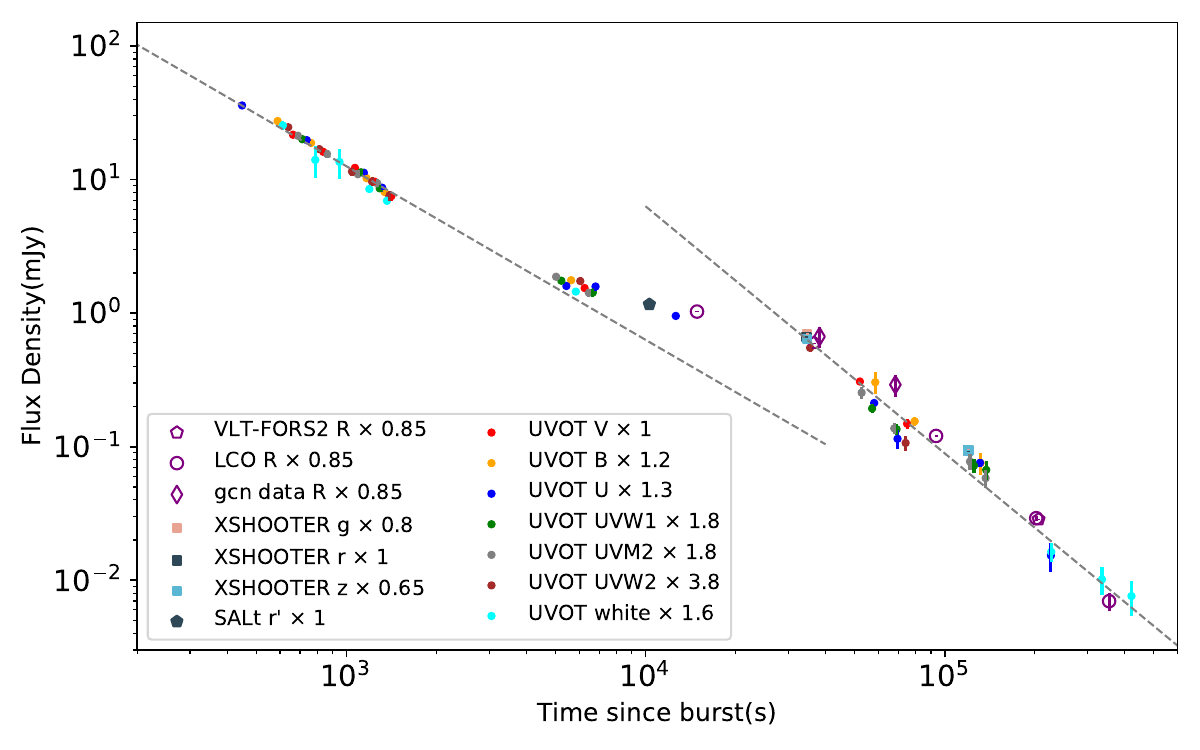}
\caption{Shifted light curves of GRB 191221B. By normalizing the flux density of each band to a comparable level, it becomes evident that even without the MASTER data, the decay index of the optical afterglow varies significantly between the early and late stages, with a shallow decay in the middle. 	\label{fig:together}}
\end{figure}

The observations from different UVOT bands exhibit rather similar behavior, with a decay index of $\alpha \approx 1.36$ before 2000 seconds. However, The light curve of the MASTER clear-band shows a decay index of $\alpha = 1.13 \pm 0.05$ before 3500 seconds. 
The bump in the X-ray light curve at about $10^3$ s is inconsistent with optical behavior, suggesting that they come from different emission regions. The optical light curve displays a plateau during phase D with a decay index of $\alpha = 0.5$, consistent across other bands excluding the MASTER clear-band data (refer to Figure \ref{fig:together} for a clearer representation of this characteristic), whereas no clear plateau is observed in the X-ray light curve, which exhibits a decay index of $\alpha_X = 0.87 \pm 0.11$ (with $\chi^2/\text{d.o.f.} = 12.46/13$) during this phase. This motivates us to seek a reasonable explanation for the observed behavior of GRB 191221B.

\begin{table}
\centering
\caption{Fitting results of the power-law function to the light curves of GRB 191221B.}
\label{table_fit}
\begin{tabular}{lccc}
\hline
Time interval & Filters & $\alpha$ & $\chi^2/\text{d.o.f}$ \\
\hline
(A) $300-1500$ s & UVOT & $1.36 \pm 0.03$ & 14.31/18\\
(B) $1500-3500$ s & clear-band & $1.13 \pm 0.05$ & 2.65/7\\
(C) $3500-10000$ s & clear-band & $0.86 \pm 0.02$ & 54.53/30\\
(D) $12000-45000$ s & clear-band & $0.50 \pm 0.01$ & 392.72/112\\
& R-band & $0.52 \pm 0.01$ & 0.52/1\\
(E) $> 45000$ s & R-band & $1.83 \pm 0.03$ & 9.76/3\\
& UVOT & $1.69 \pm 0.10$ & 13.89/8\\
\hline
\end{tabular}
\end{table}

\section{Physical implication for afterglow emission}	\label{sec:model}

\begin{figure*}
\includegraphics[width=1\textwidth]{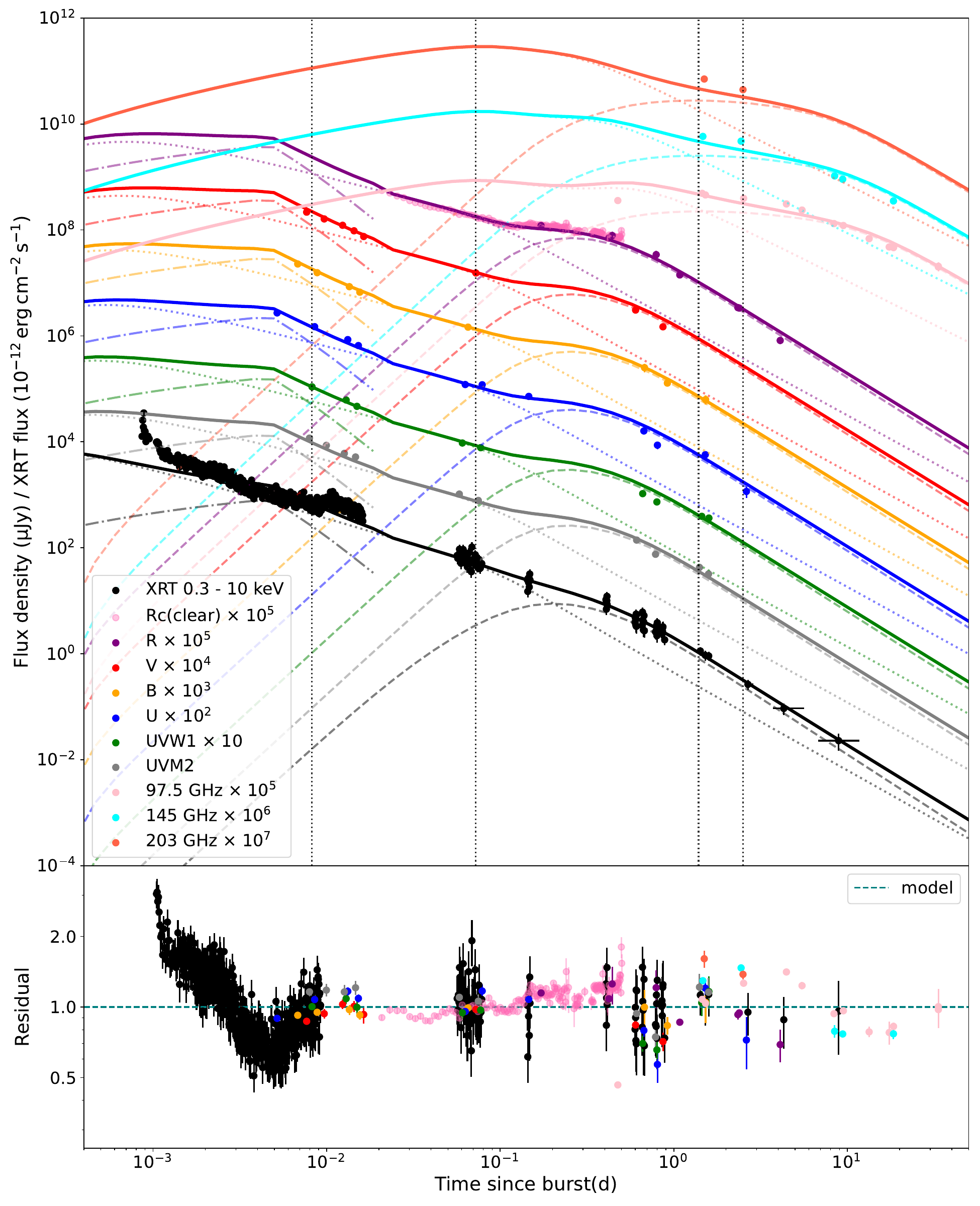}
\caption{The multiwavelength observations of GRB 191221B, and the light curves calculated from the two-component jet model. Specifically, the afterglow of the narrow and wide components, reverse shock emission, and the sum of the components were represented by dotted, dashed, dash-dotted, and solid lines, respectively. The residual in the bottom panel represent the ratio of the flux density of the observation to the model.}
\label{fig:lc_fit}
\end{figure*}

We start by considering the afterglow emission produced by the simple forward shock. The presence of an optical plateau during phase D implies that the afterglow light curve does not conform to a simple decay model (see Figure \ref{fig:together}). Instead, it exhibits characteristics indicative of energy injection.

Energy injection into the forward shock has been known to cause plateaus \citep{1998A&A...333L..87D,Panaitescu_1998,Rees_1998,Zhang_2001}, as some fraction of the energy of the burst was carried by slow-moving shells that were ejected at late times then catch up and collide with faster decelerating shells. However, such plateaus caused by energy injection should exhibit achromatic behavior and be observed in the X-ray band \citep{Kumar_2000}. The absence of an X-ray plateau in the light curve of GRB 191221B suggests that energy injection is unlikely to have played a role in this case.

Here, we propose that the optical and X-ray emissions from this burst can be explained by the two-component jet model.

\subsection{Two-component model}

The two-component jet model contains a narrow ultra-relativistic component responsible for early afterglow, and a wide, mildly relativistic component that produces the later afterglow emission \citep{Berger2003-go,Peng_2005}. 

We use a Python-Fortran hybrid code \texttt{ASGARD} package that has been used in \citet{2023ApJ...947...53R} to calculate the dynamics and radiation physics of the afterglow. The synchrotron radiation and synchrotron self-Compton (SSC) processes of electrons in the jet are considered to generate realistic radiation behavior.

The parameters are: the viewing angle $\theta_{\rm v}$, jet opening angle $\theta_{\rm j}$, initial Lorentz factor $\Gamma_0$, isotropic equivalent kinetic energy $E_{\rm K,iso}$, the spectral index of electron distribution $p$, circumburst density $n_0$, the fraction of shock energy to electron $\epsilon_{\rm e}$ and to the magnetic field $\epsilon_{\rm B}$.

The emission from the sides of the jet is relatively weak, so distant GRBs can only be observed when the symmetry axis is close to the line-of-sight direction, known as on-axis observation. To simplify, we assumed that both jets are viewed on-axis and are coaxial, with the viewing angle $\theta_{\rm v}$ set to 0, and the values of $n_0$ common for both jets.

According to our analysis, the temporal slope of UVOT observations prior to 2000 s is steeper than the slope of the MASTER clear-band between 2000 s and 4000 s (see Figure \ref{fig:lc}). This difference may be attributed to the emission of the reverse shock (RS) during the early stage. Notably, the flux density of the RS emission is higher than that of the forward shock (FS) emission and exhibits a decay index of 1.4 from about  400 s to 2000 s. However, the RS emission gradually fades below the FS emission at around 2000 s. To account for this reverse shock component, we adopt the numerical method presented by \citet{YanTing_2007} and the Fortran program they developed, and introduce three additional parameters to the model fit, namely $\epsilon_{\rm e}$, $\epsilon_{\rm B}$, and $p$ of the reverse shock. We find that the RS emission from the narrow component is significant, its isotropic kinetic energy is $E_{{\rm K, iso}, n}$, while the RS emission from the wide component is negligible.

\subsection{Numerical results}

\begin{figure}
\includegraphics[width=1\columnwidth]{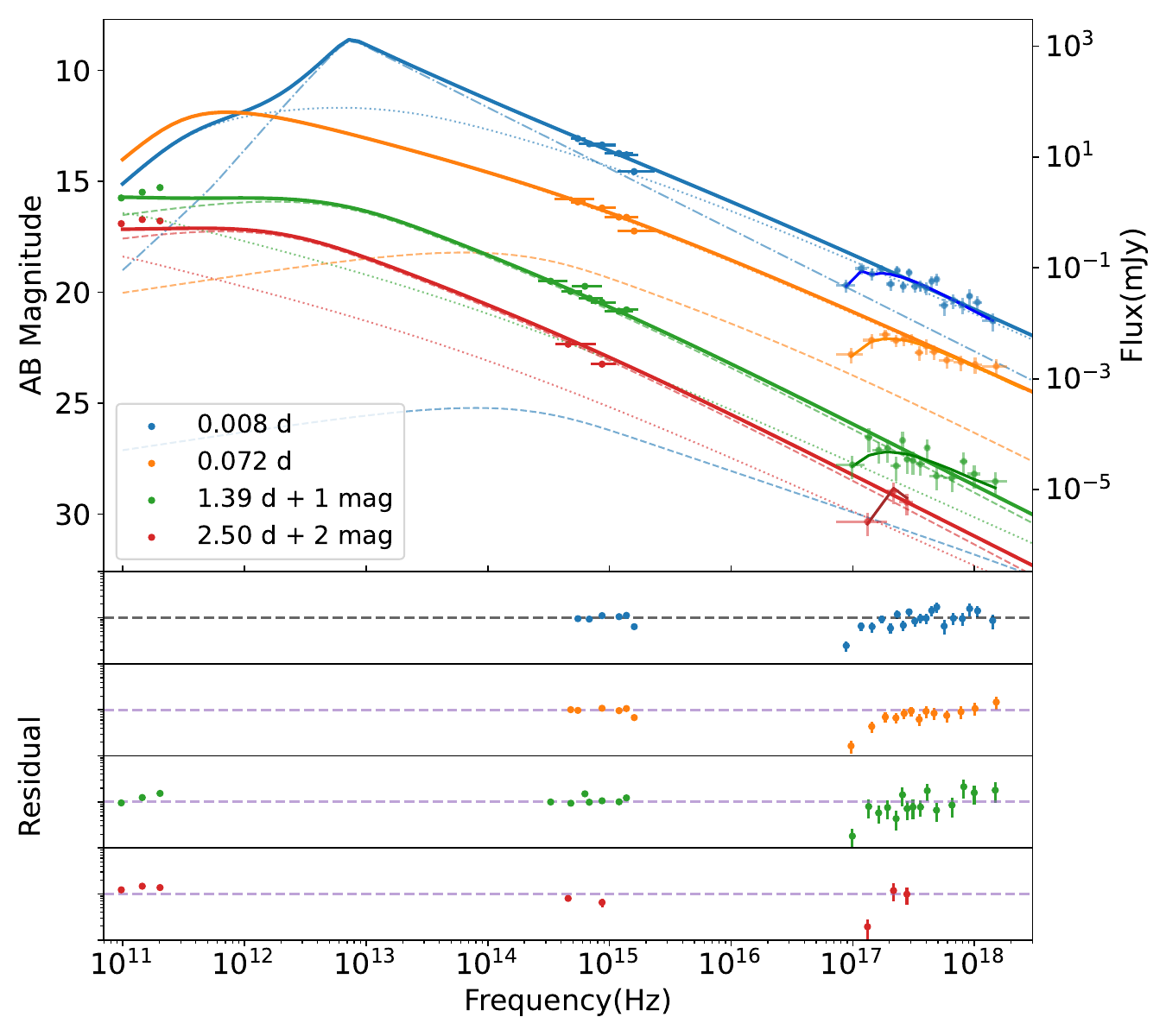}
\caption{The SED of the afterglow of GRB 191221B at the time of 710, 6230, 120000 seconds(0.008, 0.072, 1.39 days). The narrow component, wide component, early reverse shock emission, and the sum of all components were represented by dotted, dashed, dash-dotted, and solid lines, respectively. The residuals represent the ratio of the flux density of the observation to the model at these four time points. Notably, the flux in the UVW2 filter was observed to be lower than anticipated, possibly due to Lyman-alpha absorption.    \label{fig:sedall}}
\end{figure}

The results of the fitting are displayed in Figure \ref{fig:lc_fit}, where the light curves of the afterglow are shown. The model parameters are listed in Table \ref{table_par}. The parameters of the reverse shock are $\epsilon_{\rm e}=0.08$, $\epsilon_{\rm B}=0.3$, $p=2.2$. The data from the UVW2 band was not used due to its lower flux compared to the expected values from the SEDs (see Figure \ref{fig:sedall}), they may be affected by Lyman-alpha absorption. The X-ray data acquired prior to 90 seconds were obtained in the XRT-WTSLEW mode, which is a window timing mode utilized during spacecraft slew. Additionally, the bump observed at $10^3$ s was most likely a result of central activities and could not be accurately reproduced by the model, hence these data can be disregarded in the analysis.

It can be seen that the narrow jet component is responsible for the achromatic decay from $10^2$ to $10^3$ seconds, while the optical plateau and the late optical/UV/X-ray afterglow can be explained by the wide jet component. The break in the R-band light curve at 0.5-day is the jet break of the wide component. The radio afterglow is also dominated by the wide component, as shown in Figure \ref{fig:lc_fit}. However, the radio observations are not well-reproduced, possibly due to interstellar scintillation disturbances \citep{1984A&A...134..390R,1986ApJ...307..564R}.

In Figure \ref{fig:sedall} we presented the SEDs of the afterglow of GRB 191221B at 0.008-day, 0.072-day, 1.39-day, and 2.50-day. These four times correspond to the four vertical black dotted lines in Figure \ref{fig:lc_fit}. 
Additionally, Figure \ref{fig:sedall} shows the spectra of our two-component jet model at four times with blue, orange, green and red lines, respectively.
The components of the model provide a good fit to the observed data. It is worth noting that at 0.008 days, the contribution of the RS emission in the optical band is significant. According to the model, the optical to X-ray spectral indices at the four time points are $\beta=(0.94,0.90,1.06,1.06)$. By excluding the UVW2 band data affected by Lyman-alpha absorption, the low-energy X-ray data (around $10^{17}$ Hz) affected by hydrogen absorption, and radio data influenced by interstellar scintillation disturbances, we calculate the reduced chi-square of the modeled SEDs to be $\chi^2/\text{d.o.f.} = 129.84/ 50$.

As found by \citet{Urata2022-191221}, the spectral slope of the radio afterglow changed from $\beta=-0.3$ at 2.5 days to $\beta=0.7$ at 9.5 and 18.4 days, implying that the injection frequency of the emission $\nu_m$ crossed the radio band at around $t \approx 4$ days, and the electron spectral slope $p=2.4$ can be inferred from the relation $\beta = (p-1)/2$ for $\nu_m< \nu < \nu_c$ case \citep{zhang2004gamma}. It is in agreement with our fitting result of $p=2.47$ for the wide component.

If we directly fit the optical to X-ray spectral data with a single power-law function, the index is found to be 0.95 at 0.008-day and 1.05 at 1.39-day. So for the wide component, the ratio of the X-ray flux to the optical flux is smaller than that of the narrow component, which is the reason why the plateau is more pronounced in optical bands than in the X-ray.

\begin{table}
\centering
\caption{Parameters of GRB 191221B afterglow assuming a two-component jet model.}
\label{table_par}
\begin{tabular}{ccc}
\hline
    & Narrow jet & Wide jet \\
\hline
$\theta_{\rm j}$ (rad) & $0.025$ & $0.048$\\
$\log_{10}(\Gamma_0)$ & $2.6$ & $1.4$ \\
$p$ & $2.04$ & $2.47$\\
$\log_{10}(E_{\rm K,iso})$ & $54.40$ & $53.99$\\
$\log_{10}(n_0)$ & \multicolumn{2}{c}{$0.713$}\\
$\log_{10}(\epsilon_{\rm e})$ & $-1.80$& $-0.69$\\
$\log_{10}(\epsilon_{\rm B})$ & $-2.76$& $-3.77$\\
\hline
\end{tabular}
\end{table}

\section{discussion}	\label{sec:dis}

The flattening of the optical light curve in the afterglow of GRB 191221B has been previously reported by \citet{Buckley_2021}. The MASTER photometric data were acquired using an apparent bandwidth (unfiltered) rather than the standard bandwidth.
Then we used our usual procedure \citep{2017MNRAS.465.3656L} for calibrating stellar values repeatedly tested/described earlier \citep{2008ApJ...685..376K}.
Extensive experience with our clear filter, calculated using the formula 0.2B + 0.8R, indicates that it is closest to the standard Rc-filter.
Naturally, precise calibration relies on the source's spectrum, which is not required to be known beforehand. Fortunately, MASTER's wide-field images enable the utilization of numerous reference stars with varying colors and magnitudes, all proximate to the afterglow's brightness at each instance. This evidence of a plateau's presence within the light curve is established through a comparative analysis involving reference stars possessing roughly equivalent stellar magnitudes.

In our study, we suggest that the multiband afterglow of GRB 191221B can be explained by a two-component jet model. We reduced the image data obtained from Swift/UVOT, VLT, and LCO and used a two-component jet model to calculate the multi-wavelength light curve of the GRB 191221B afterglow. The best-fitting result of the model is presented in Figure \ref{fig:lc_fit}. This model features a narrow jet ($\theta_j = 0.025$ rad / $1.4^{\circ}$) with a high initial Lorentz factor ($\Gamma_0 = 400$) and a wider ($\theta_j = 0.048$ rad / $2.8^{\circ}$), but slower ($\Gamma_0 = 25$) jet that surrounds the narrow component. 

The formation of the two components may occur when neutrons and protons decouple in a neutron-rich, hydro-dynamically driven jet that originates from a neutron-rich accretion disc of a collapsed massive star, as described by \citet{Vlahakis_2003}.
The ratio of the kinetic energy $E_{\rm K}$ and the isotropic equivalent kinetic energy $E_{\rm K,iso}$ is given by the beaming factor $f_b = 1 - \cos \theta_j \approx \theta_j^2 / 2$, which allows us to calculate the kinetic energy of the two components: $E_{{\rm K},n} = 7.9 \times 10^{50} \rm erg$ and $E_{{\rm K},w} = 1.1 \times 10^{51} \rm erg$. According to \citet{Peng_2005}, within an initially neutron-rich, hydromagnetically accelerated jet, the ratio $E_{{\rm K},w} / E_{{\rm K},n}>1$ and $E_{\text{K,iso}, w} / E_{\text{K,iso}, n} < 1$ satisfies the condition that the narrow component dominates the early afterglow and the wide component can dominate at late times.

The two-component jet case is rarely seen in the GRB afterglows but this model provides an interpretation for the chromatic rebrightening in the GRB light curves. And the discovery of a gamma-ray burst with two-component jets has implications for our understanding of the GRB population. In the analysis of GRB 191221B, we assumed that the viewing angle $\theta_{\rm v} = 0$, because GRB with such a high-redshift burst (z = 1.148) can only be detected at a very small viewing angle. For analysis and fitting of nearby GRBs, the viewing angle distribution and the jet structure may be better constrained.

\section*{Acknowledgements}

This work was supported in part by NSFC under grants of No. 11933010, 11921003, 12073080, 12233011 and 12225305, Key Research Program of Frontier Sciences (No. QYZDJ-SSW-SYS024), the China Manned Space Project (NO.CMS-CSST-2021-A13), Major Science and Technology Project of Qinghai Province (2019-ZJ-A10).

\section*{Data Availability}

The VLT, LCO and Swift data analysed in this work are all publicly available. The MASTER data analysed in this work will be shared on reasonable request to the MASTER (VL). Some MASTER data (light curves and spectra) are available at the following website: \url{https://tinyurl.com/yyd3hty8}. Other data underlying this article will be shared on reasonable request to the corresponding authors.



\bibliographystyle{mnras}
\bibliography{ref} 




\appendix

\section{Additional Reference Stars in Calibration of MASTER Data}
\onecolumn

\begin{figure*}
\includegraphics[width=1\textwidth]{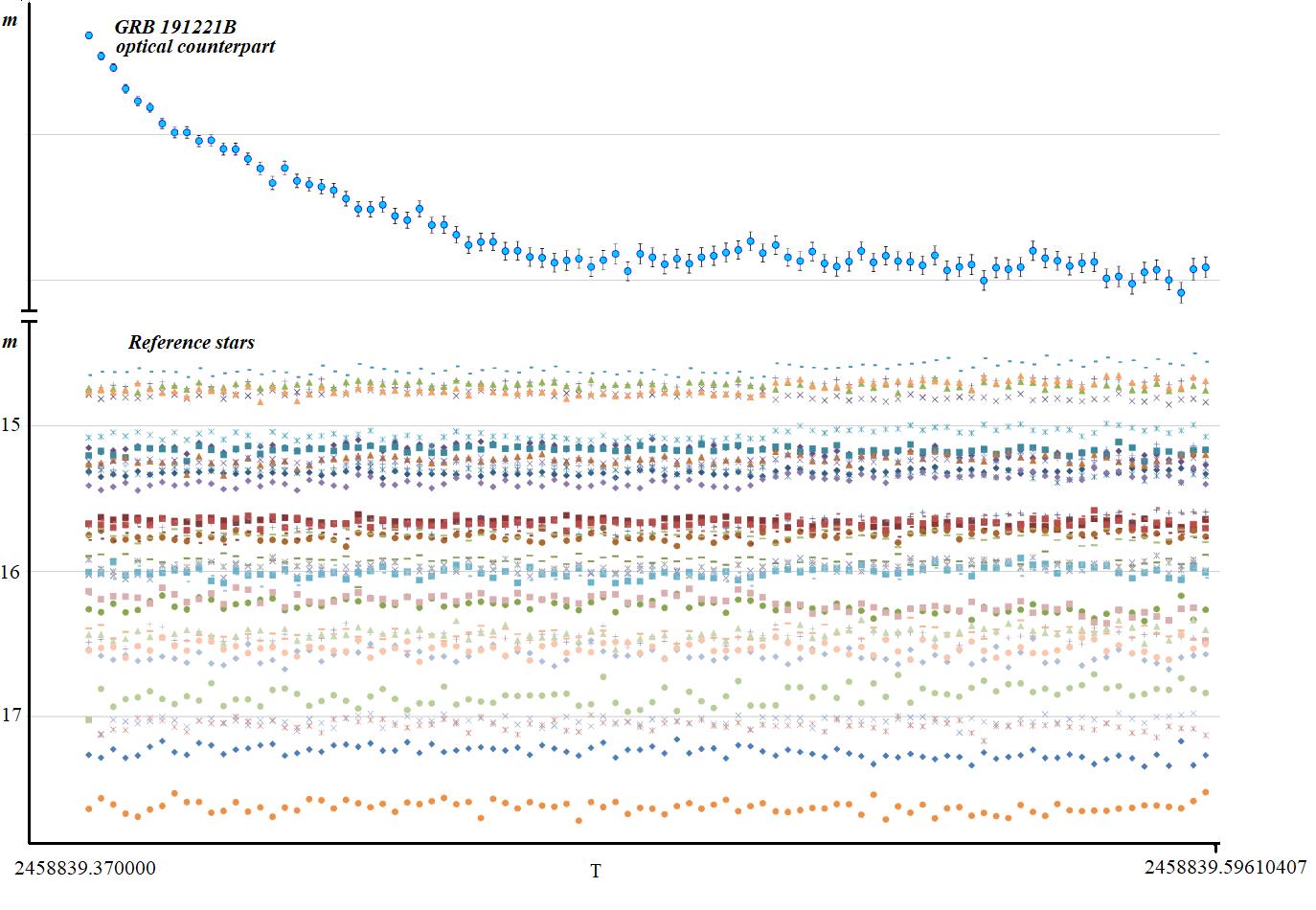}
\caption{Photometry results for the comparison stars in each exposure image of MASTER observation.   \label{fig:refstars}}
\end{figure*}

\section{Observation log and photometry for GRB 191221B}

\begin{longtable}{cccccc}
        \caption{Observation log and photometry for GRB 191221B.}
        \label{table_191221B} \\
    \hline
    Time since GRB & Time since GRB & Exposure time & Instrument & Filter & Magnitude AB \\
    (seconds) & (days) & (s) &  &  & \\
    \hline
    \endfirsthead

    \hline
    Time since GRB & Time since GRB & Exposure time & Instrument & Filter & Magnitude AB \\
    (seconds) & (days) & (s) &  &  & \\
    \hline
    \endhead
    
    \hline
    \multicolumn{6}{r}{\textit{Continued on next page}}
    \endfoot
    
    \hline
    \endlastfoot
    
98     & 0.0011 & 9.2    & SWIFT/UVOT & V     & 11.66 ± 0.13           \\
663    & 0.0077 & 19.4   & SWIFT/UVOT & V     & 13.15 ± 0.04           \\
837    & 0.0097 & 19.5   & SWIFT/UVOT & V     & 13.47 ± 0.05           \\
1068   & 0.0124 & 19.4   & SWIFT/UVOT & V     & 13.77 ± 0.05           \\
1243   & 0.0144 & 19.5   & SWIFT/UVOT & V     & 14.03 ± 0.06           \\
1413   & 0.0164 & 12.7   & SWIFT/UVOT & V     & 14.30 ± 0.09           \\
6256   & 0.0724 & 196.6  & SWIFT/UVOT & V     & 15.97 ± 0.05           \\
52058  & 0.6025 & 885.1  & SWIFT/UVOT & V     & 17.67 ± 0.06           \\
74864  & 0.8665 & 844.5  & SWIFT/UVOT & V     & 18.29 ± 0.09           \\
227424 & 2.6322 & 645.25 & SWIFT/UVOT & V     & > 19.83 \\
\\
589    & 0.0068 & 19.4   & SWIFT/UVOT & B     & 13.12 ± 0.04           \\
762    & 0.0088 & 19.4   & SWIFT/UVOT & B     & 13.53 ± 0.04           \\
1168   & 0.0135 & 19.4   & SWIFT/UVOT & B     & 14.19 ± 0.05           \\
1342   & 0.0155 & 19.4   & SWIFT/UVOT & B     & 14.45 ± 0.04           \\
5640   & 0.0653 & 196.6  & SWIFT/UVOT & B     & 16.1 ± 0.03            \\
58590  & 0.6781 & 53.70  & SWIFT/UVOT & B     & 17.93 ± 0.21           \\
79362  & 0.9185 & 380.4  & SWIFT/UVOT & B     & 18.53 ± 0.08           \\
131823 & 1.5257 & 211.3  & SWIFT/UVOT & B     & 19.28 ± 0.19           \\
226775 & 2.6247 & 622.6  & SWIFT/UVOT & B     &  > 20.85 \\
\\
448    & 0.0052 & 245.8  & SWIFT/UVOT & U     & 12.97 ± 0.03           \\
737    & 0.0085 & 19.4   & SWIFT/UVOT & U     & 13.62 ± 0.03           \\
1144   & 0.0132 & 19.4   & SWIFT/UVOT & U     & 14.23 ± 0.04           \\
1317   & 0.0152 & 19.4   & SWIFT/UVOT & U     & 14.51 ± 0.04           \\
5435   & 0.0629 & 196.6  & SWIFT/UVOT & U     & 16.29 ± 0.04           \\
6805   & 0.0788 & 65.4   & SWIFT/UVOT & U     & 16.35 ± 0.06           \\
12623  & 0.1461 & 505.1  & SWIFT/UVOT & U     & 16.80 ± 0.04           \\
58104  & 0.6725 & 885.1  & SWIFT/UVOT & U     & 18.40 ± 0.06           \\
69551  & 0.8050 & 90.2   & SWIFT/UVOT & U     & 18.95 ± 0.20           \\
131248 & 1.5191 & 885.12 & SWIFT/UVOT & U     & 19.66 ± 0.07           \\
226454 & 2.6210 & 622.55 & SWIFT/UVOT & U     & 21.04 ± 0.24           \\
334900 & 3.8762 & 596.0  & SWIFT/UVOT & U     & > 21.63 \\
\\
712    & 0.0082 & 19.4   & SWIFT/UVOT & UVW1  & 14.03 ± 0.04           \\
1119   & 0.0130 & 19.5   & SWIFT/UVOT & UVW1  & 14.65 ± 0.05           \\
1292   & 0.0150 & 19.4   & SWIFT/UVOT & UVW1  & 14.95 ± 0.06           \\
5230   & 0.0605 & 196.6  & SWIFT/UVOT & UVW1  & 16.64 ± 0.05           \\
6666   & 0.0772 & 196.6  & SWIFT/UVOT & UVW1  & 16.84 ± 0.05           \\
57194  & 0.6620 & 885.6  & SWIFT/UVOT & UVW1  & 19.01 ± 0.07           \\
69048  & 0.7992 & 885.6  & SWIFT/UVOT & UVW1  & 19.35 ± 0.08           \\
125238 & 1.4495 & 669.6  & SWIFT/UVOT & UVW1  & 20.01 ± 0.13           \\
137657 & 1.5933 & 469.1  & SWIFT/UVOT & UVW1  & 20.13 ± 0.16           \\
244988 & 2.8355 & 740.5  & SWIFT/UVOT & UVW1  & > 21.12 \\
\\
688    & 0.0080 & 19.4   & SWIFT/UVOT & UVM2  & 13.98 ± 0.05           \\
861    & 0.0100 & 19.4   & SWIFT/UVOT & UVM2  & 14.32 ± 0.06           \\
1093   & 0.0127 & 19.4   & SWIFT/UVOT & UVM2  & 14.70 ± 0.07           \\
1268   & 0.0147 & 19.4   & SWIFT/UVOT & UVM2  & 14.87 ± 0.07           \\
5024   & 0.0581 & 196.6  & SWIFT/UVOT & UVM2  & 16.62 ± 0.06           \\
6461   & 0.0748 & 196.6  & SWIFT/UVOT & UVM2  & 16.92 ± 0.07           \\
52769  & 0.6108 & 497.2  & SWIFT/UVOT & UVM2  & 18.79 ± 0.10           \\
67841  & 0.7852 & 885.58 & SWIFT/UVOT & UVM2  & 19.46 ± 0.10           \\
121470 & 1.4059 & 769.8  & SWIFT/UVOT & UVM2  & 20.00 ± 0.15           \\
137022 & 1.5859 & 885.6  & SWIFT/UVOT & UVM2  & 20.25 ± 0.15           \\
244607 & 2.8311 & 728.9  & SWIFT/UVOT & UVM2  & > 21.24 \\
\\
639    & 0.0074 & 19.4   & SWIFT/UVOT & UVW2  & 14.63 ± 0.05           \\
812    & 0.0094 & 19.4   & SWIFT/UVOT & UVW2  & 15.04 ± 0.06           \\
1044   & 0.0121 & 19.4   & SWIFT/UVOT & UVW2  & 15.46 ± 0.07           \\
1219   & 0.0141 & 19.4   & SWIFT/UVOT & UVW2  & 15.65 ± 0.08           \\
1391   & 0.0161 & 19.4   & SWIFT/UVOT & UVW2  & 15.91 ± 0.09           \\
6051   & 0.0700 & 196.6  & SWIFT/UVOT & UVW2  & 17.51 ± 0.07           \\
35475  & 0.4106 & 672.0  & SWIFT/UVOT & UVW2  & 18.76 ± 0.07           \\
74035  & 0.8569 & 885.6  & SWIFT/UVOT & UVW2  & 20.54 ± 0.14           \\
244201 & 2.8264 & 728.8  & SWIFT/UVOT & UVW2  & > 21.64 \\
\\
185    & 0.0021 & 147.4  & SWIFT/UVOT & white & 12.02 ± 0.04\footnote{
The saturated source in the first exposure of white filter is too bright and exceeds the correction range. The correction applied to it was an extrapolation beyond the range studied by \citet{2023NatAs.tmp..135J}, and it is uncertain whether it is reliable.}    \\
613    & 0.0071 & 19.4   & SWIFT/UVOT & white & 13.39 ± 0.06    \\
787    & 0.0091 & 19.4   & SWIFT/UVOT & white & 14.03 ± 0.26    \\
948    & 0.0110 & 147.4  & SWIFT/UVOT & white & 14.07 ± 0.25    \\
1193   & 0.0138 & 19.4   & SWIFT/UVOT & white & 14.71 ± 0.04           \\
1366   & 0.0158 & 19.4   & SWIFT/UVOT & white & 14.93 ± 0.04           \\
5845   & 0.0677 & 196.6  & SWIFT/UVOT & white & 16.63 ± 0.05           \\
227070 & 2.6281 & 622.6  & SWIFT/UVOT & white & 21.50 ± 0.16           \\
335515 & 3.8833 & 596.0  & SWIFT/UVOT & white & 22.02 ± 0.25           \\
421087 & 4.8737 & 524.91 & SWIFT/UVOT & white & 22.33 ± 0.32           \\
476082 & 5.5102 & 244.86 & SWIFT/UVOT & white & > 22.09\\
\\
36607 & 0.4237 & 30 × 1 & VLT/FORS2 & R\_SPECIAL & 16.854 ± 0.002\\
206001 & 2.3843 & 30 × 2 & VLT/FORS2 & R\_SPECIAL & 20.159 ± 0.008\\
\\
34477 & 0.3990 & 15 × 3 & VLT/X-SHOOTER & r\_prime & 16.91 ± 0.02 \\
34586 & 0.4003 & 20 × 3 & VLT/X-SHOOTER & z\_prime & 16.47 ± 0.01 \\
34694 & 0.4016 & 15 × 3 & VLT/X-SHOOTER & g\_prime & 16.67 ± 0.02 \\
119821 & 1.3868 & 15 × 3 & VLT/X-SHOOTER & r\_prime & 19.04 ± 0.02 \\
119929 & 1.3881 & 20 × 3 & VLT/X-SHOOTER & z\_prime & 18.55 ± 0.01 \\
120035 & 1.3893 & 15 × 3 & VLT/X-SHOOTER & g\_prime & 18.83 ± 0.02 \\
\\
10301 & 0.1192 & 5.0 & SALTICAM & SDSSr-S1 & 16.30 ± 0.01 \\
\\
14859 & 0.1720 & 90 × 10 & LCO/fa06 & R & 16.27 ± 0.01 \\
93569 & 1.0830 & 240 × 1 & LCO/fa06 & R & 18.60 ± 0.02 \\
202246 & 2.3408 & 300 × 2 & LCO/fa03 & R & 20.14 ± 0.05 \\
354488 & 4.1029 & 300 × 3 & LCO/fa16 & R & 21.69 ± 0.17 \\
\end{longtable}


\bsp	
\label{lastpage}
\end{document}